\documentclass[a4paper,11pt]{article}
\pdfoutput=1 

\usepackage{jinstpub} 
\usepackage{upgreek}
\usepackage{cancel}
\usepackage{nicefrac}
\usepackage{graphicx}
\usepackage{subcaption}
\usepackage{lineno}
\usepackage{indentfirst}
\usepackage{svg}

\title{\boldmath Low-field carrier mobilities in silicon irradiated to extreme fluences}

\author[b]{I. Bloch,}
\author[b]{B. Bruers,}
\author[c]{C.-T. Klein,}
\author[a]{H. Lacker,}
\author[a]{P. Li,}
\author[d]{M. Ullan,}
\author[e]{Y. Unno,}
\author[f]{I. Mandi\'c,}
\author[a,1]{C. Scharf\note{Corresponding author.}}

\affiliation[a]{Humboldt University of Berlin, Unter den Linden 6, 10117 Berlin, Germany}
\affiliation[b]{Deutsches Elektronen-Synchrotron (DESY), Platanenallee 6, 15738 Zeuthen, Germany}
\affiliation[c]{Carleton University, 1125 Colonel By Dr., Ottawa, ON, K1S 5B6, Canada}
\affiliation[d]{Instituto de Microelectr\'onica de Barcelona (IMB-CNM), CSIC, 
08193 Barcelona, Spain}
\affiliation[e]{Institute of Particle and Nuclear Studies, High Energy Accelerator Research Organization (KEK), 1-1 Oho, Tsukuba, Ibaraki 305-0801, Japan} 
\affiliation[f]{Jo\v{z}ef Stefan Institute, Jamova 39, SI-1000 Ljubljana, Slovenia} 

\emailAdd{scharfch@hu-berlin.de}

\abstract{ 
The low-field carrier mobilities in <100> silicon were quantified as a function of the 1\,MeV neutron-equivalent fluence up to $10^{18}\,$cm$^{-2}$ and for temperatures between 230\,K and 260\,K. 
Current measurements were fitted using a mobility model for scattering at ionized impurities. Technology-aided design (TCAD) simulations were compared to measurements and used to estimate the carrier concentrations, which are parameters in the fit. The fit model describes the data very well, both as a function of fluence and the temperature. At a fluence of $6 \cdot 10^{17}\,$cm$^{-2}$, which is expected for the innermost detector layers at the proposed Future Circular Hadron Collider (FCC-hh), the electron mobility was found to decrease by $\sim60$\% and the hole mobility by $\sim45$\%.
}


\proceeding{21$^{\text{th}}$ Anniversary "Trento" Workshop on Advanced Silicon Radiation Detectors\\  Feb. 2026, Perugia}

\begin{document}
\maketitle
\flushbottom

\section{Introduction}
\label{sec:intro}

We present a study of the forward and reverse currents in silicon pad diodes irradiated to extreme neutron fluences of up to $\Phi_{\mathrm{eq}} = 10^{18}\,$cm$^{-2}$, slightly exceeding the expected fluences at the innermost radii of tracking detectors at the proposed Future Circular Hadron Collider (FCC-hh)\,\cite{fcc}. At such extreme fluences, the low-doped silicon bulk and the highly doped implant no longer behave as a conventional pn diode\,\cite{thesis}. In equilibrium, excess free carriers are trapped at radiation-induced deep defects, compensating ionized shallow defects in the bulk.
The Fermi level becomes pinned near mid-gap, reducing the bulk carrier concentrations to near-intrinsic levels and significantly increasing the resistivity of the bulk\,\cite{bulkcompsim}. 
Coulomb scattering at ionized defects\,\cite{long59} is therefore expected to reduce the low-field carrier mobilities in silicon irradiated at extreme fluences.
In contrast, high-field velocity saturation is largely unaffected by Coulomb scattering\,\cite{jacoboni}. The fluence-dependence of the mobilities is required for simulation of detectors at future hadron colliders. At the time of writing, few mobility measurements were published, mostly for fluences of $\lesssim10^{16}\,$cm$^{-2}$ where the reduction is still small\,\cite{fretwurst, thesis}.

To quantify the mobility degradation caused by radiation-induced ionized impurity scattering as a function of fluence and to obtain a qualitative understanding of the diode’s behavior, current-voltage characteristics of irradiated diodes were measured at various temperatures. These measurements are compared to TCAD simulations using different radiation-damage models to evaluate their ability in reproducing the observed currents at such extreme fluences, and to gain insight into the underlying physical processes driving these changes.

\section{Experimental methods}

\subsection*{Measurements}
For this study, planar "MD8" diodes from wafers of the ATLAS Inner Tracker (ITk) silicon strip sensor submission ATLAS18\,\cite{itk} were investigated. The MD8 diodes are about $8 \cdot 8$\,mm$^2$ and feature a large pad surrounded by a guard ring (GR). The bulk of the diodes has an active thickness of about $d=295$\,$\upmu$m and a p-type doping concentration before irradiation of $4.2\cdot10^{12}$\,cm$^{-3}$. The pn junction is located at the top and the backside contact is highly p-doped. The diodes were irradiated with neutrons to fluences between $5 \cdot 10^{16}$ and $10^{18}\,$cm$^{-2}$ at the research reactor of JSI, Ljubljana\,\cite{jsi}. The uncertainty on the fluence is assumed to be 10\%.

The measurements were performed in a temperature-controlled probe station in a dry air environment, with a dew point well below the set temperature. The diodes were attached to copper plates using conductive silver paste, to establish electrical and thermal contact. The pad and the GR were wire-bonded to PCBs and connected to ground via two Keithley 6485 pico-ampere meters. Voltages between $U=\pm1000$\,V were applied to the backside contact using a Keithley 2410. 

The diode current is extremely sensitive to the temperature, and self-heating in irradiated diodes due to their high current can distort the local temperature. Therefore, a temperature sensor was attached close to the diode on the copper plate using thermal pads to enable precise temperature control. The temperature was measured using PT1000 temperature sensors read out by a Keithley 2001. The temperature can be controlled precisely down to about 235 K and the dry air is pre-cooled 
for temperature stability. The temperature was stable within $\pm0.1$\,K during measurements.

\subsection*{Simulations}
In order to gain a better understanding of the measurements, simulations were performed with Synopsys TCAD\,\cite{tcad}. The sensor geometry in the simulations is a 2D cross section through the pad region of the diode, using the dimensions and bulk doping mentioned earlier.
The absolute peak doping concentrations of the implants are $10^{19}$\,cm$^{-3}$. 

Surface currents from the silicon bulk-oxide interface were assumed to be negligible after extreme fluences. The positive oxide charge and the interface trap densities are driven by ionizing radiation and less relevant for neutron irradiation. In addition, they are limited to the immediate interface and saturate at some point. The effect of the positive oxide charge will be compensated by immobile negatively charged bulk defects rather than free charge carriers after high fluences\,\cite{sioqcomp} and no electron inversion layer can form. Accordingly, only small fractions of the interface near the space charge region (SCR) at the front implant can contribute to the current and the total current is assumed to be dominated by the bulk. Geometrical edge effects should be mostly eliminated by the large pad area compared to the thickness and the use of the GR. The measured GR current was $\sim10\%$ of the pad current after irradiation, which roughly aligns with the geometrical expectations.

The effects of radiation damage were approximated in the simulation by introducing a few effective point defect energy levels, representative for the plethora of real defects. The Hamburg Penta-Trap Model\,\cite{hptm} (HPTM) with five and the Perugia model\,\cite{perugia} with three defect energy levels were compared. These damage models were optimized to approximate measurements performed for much lower fluences $\lesssim10^{16}$\,cm$^{-2}$ and are not expected to describe the extreme fluences of this work very well. The HPTM was modified by activating the setting \verb|Add2TotalDoping(ChargedTraps)|. This setting dynamically calculates the fraction of ionized defects, which is then used to calculate the mobility, but not band-gap narrowing\,\cite{bgn}. The Perugia model is using \verb|Add2TotalDoping| by default, which uses the absolute concentration defects for both ionized-defect scattering and for band-gap narrowing, also the fraction of neutral defects.

The low-field mobilities were modeled with the empirical Masetti parameterization\,\cite{masetti}. This model was developed to describe the low-field mobilities in silicon as a function of the ionized impurity concentration at $T=300$\,K. The temperature and electric field dependence were modeled with the Canali\,\cite{canali} parameterization. For the band gap energy and carrier concentrations the Slotboom model\,\cite{slotboom} was used and the van Overstraeten-de Man model\,\cite{vanover} for impact ionization.

\section{Results}

\subsection*{Current and resistivity}

\begin{figure}[htbp]
\centering 
\begin{subfigure}{0.48\textwidth}
  \centering
  \includegraphics[width=\textwidth]{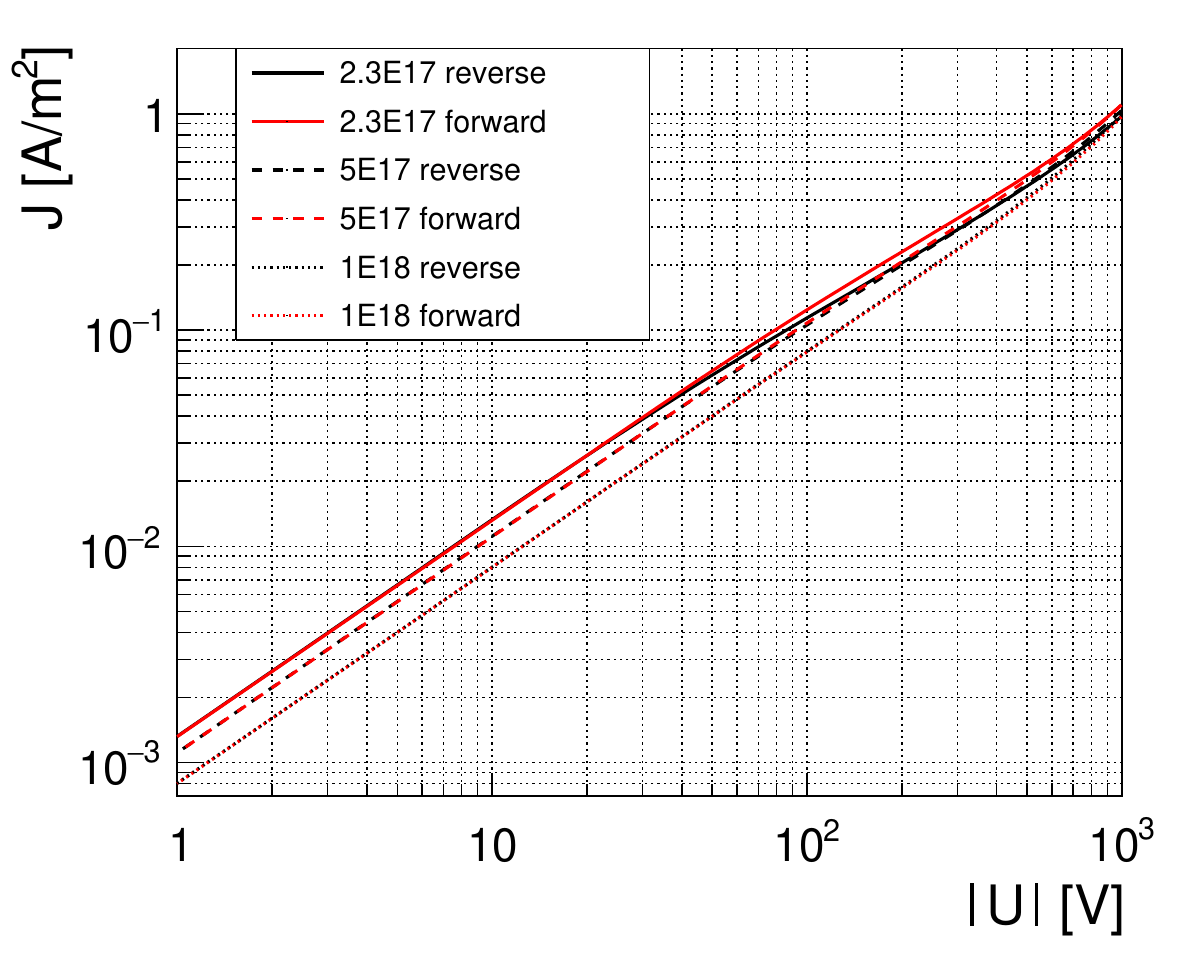}
  \caption{Current density.}
  \label{fig:current}
\end{subfigure}\hfill
\begin{subfigure}{0.48\textwidth}
  \centering
  \includegraphics[width=\textwidth]{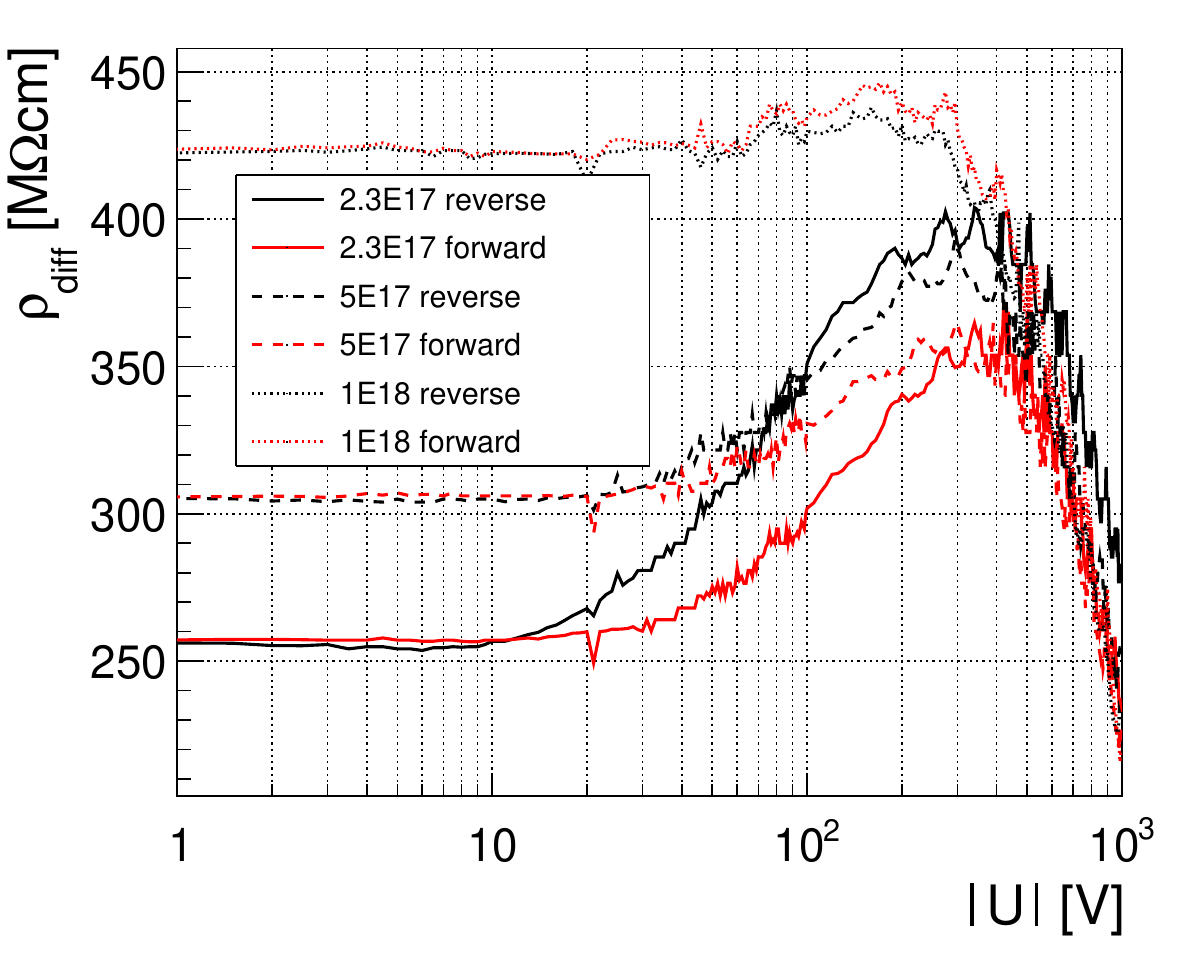}
  \caption{Differential resistivity.}
  \label{fig:resistivity}
\end{subfigure}

\caption{\label{fig:rho} (a) Measured current densities of diodes irradiated to fluences of $2.3\cdot10^{17}\,$cm$^{-2}$, $5\cdot10^{17}\,$cm$^{-2}$, and $10^{18}\,$cm$^{-2}$ at $T=236.6$\,K. Reverse bias is shown in black and forward in red, mostly overlapping. Figure\,(b) shows the resistivity calculated from Figure\,(a). For $U\lesssim10$\,V (ohmic region), the resistivity is constant and increases with fluence, while the current decreases with fluence due to the reduction in low-field mobility.  
}
\end{figure}

After irradiation with extreme particle fluences, the current density $J$ of silicon pn diodes is very similar\,\cite{thesis} for forward and reverse bias. 
As shown in Figure\,\ref{fig:rho}, for $U \lesssim 10$~V the diode behaves like a resistor, with a resistivity similar to that of intrinsic silicon\,\cite{bulkcompsim}. The differential resistivity $\rho_{\mathrm{diff}}=\frac{1}{d}\frac{\Delta U}{\Delta J}$ at these bias voltages increases with the fluence, the diode current decreases with increasing fluence. This is opposite to what is observed at lower fluences, where the reverse current increases with the fluence due to increased generation in the SCR.

For reverse bias, at a certain threshold voltage which increases with the fluence and the squared active thickness of the diode\,\cite{thesis}, the current transitions into a sublinear regime. Here, the resistivities of the thin SCRs at the highly-doped front and back implants ("double junction"\,\cite{doublej}) become significantly higher than the resistivity of the neutral bulk. This seems to increase the overall resistivity of the diode, similar to depletion in non-irradiated junctions. 

For $U\gtrsim 400$\,V the current starts to increase exponentially. This increase might be caused by field-enhanced generation\,\cite{schenk}. The measured voltage dependence of the exponential current increase does not seem to depend on the temperature in the narrow investigated range. Unfortunately, for $T>250$\,K it was not possible to apply high enough bias voltages due to the high diode current and further measurements for lower temperatures are needed to understand the temperature dependence of the current at high voltages.

For forward bias the shape of the differential resistivity is very similar to reverse bias. The forward current at very high bias voltages is slightly higher than the reverse current for all investigated fluences but for the highest fluence of $10^{18}$\,cm$^{-2}$, where the trend seems to shift.

\subsection*{Simulations}

\begin{figure}[htbp]
\centering 
\includegraphics[width=0.8\textwidth ]{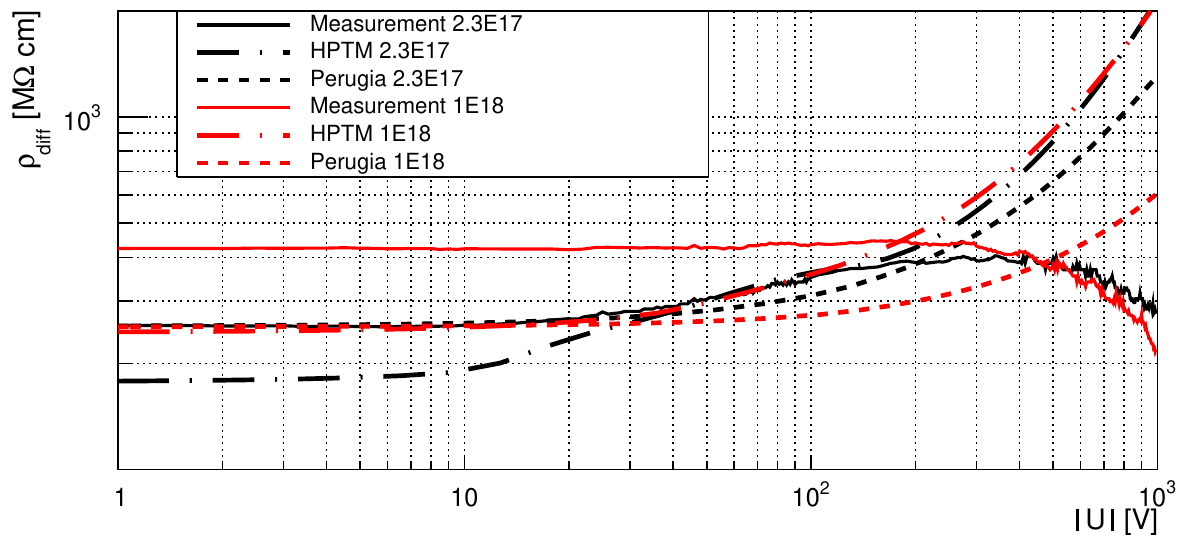}
\caption{\label{fig:meas-sim} 
Measured and simulated reverse bias differential resistivities at T = 236.6 K for two different fluences of $2.3\cdot10^{17}\,$cm$^{-2}$ (black) and $10^{18}\,$cm$^{-2}$ (red). The HPTM simulation shows increasing ohmic resistivity with the fluence like the measurements, while the Perugia simulation changes little.
}
\end{figure}


Figure\,\ref{fig:meas-sim} shows a comparison of the simulations and the measurements. In general, the investigated damage models do not reproduce the voltage dependence of the resistivities well. However, for $U\lesssim100$\,V the voltage dependence is reproduced to some degree and only at high bias voltages $U>100$\,V the deviations become large. 
The ohmic behavior for $U\lesssim 10$\,V is qualitatively reproduced by both investigated models.
The position of the transition to the sublinear regime, approximately between 10\,V and 100\,V, is not well reproduced. 
The simulated resistivities increase monotonically for high bias voltages, while the measurements decrease for $U\gtrsim400$\,V.

In the simulation, the position of the pinned Fermi level and the carrier concentrations in the bulk are determined by the implemented defect levels. 
In the HPTM simulations, the ratio of the electron $n$ and hole concentration $p$ in the compensated bulk in steady state is $\nicefrac{p}{n}=4.0$, independent of the fluence. The condition $np=n_\mathrm{i}^2=const$ is always fulfilled for the HPTM model, with the intrinsic carrier density $n_\mathrm{i}$. For the Perugia model $n_i$ increases for extreme fluences due to band-gap narrowing. 
The increase of the resistivity due to the reduction of the mobilities is observed in the HPTM simulation. The Perugia simulation does not reproduce this trend of the measurements since band-gap narrowing compensates the effect from the reduced mobility. The simulations with both models fail to reproduce the absolute resistivity values.

The mismatch at high bias voltages is not understood. Impact ionization is not significantly contributing to the reverse current in the simulations, matching the measured temperature dependence described above. 
Poole-Frenkel field-enhanced emission was tested in TCAD, but was deactivated for the results presented here (see also\,\cite{schenk}). Trap-assisted tunneling was used only for one defect for HPTM (see\,\cite{hptm}) and does not reproduce the increase. Further studies are needed understand this regime. 

\subsection*{Low-field mobility}

\begin{figure}[htbp]
\centering 
\includegraphics[width=0.8\textwidth ]{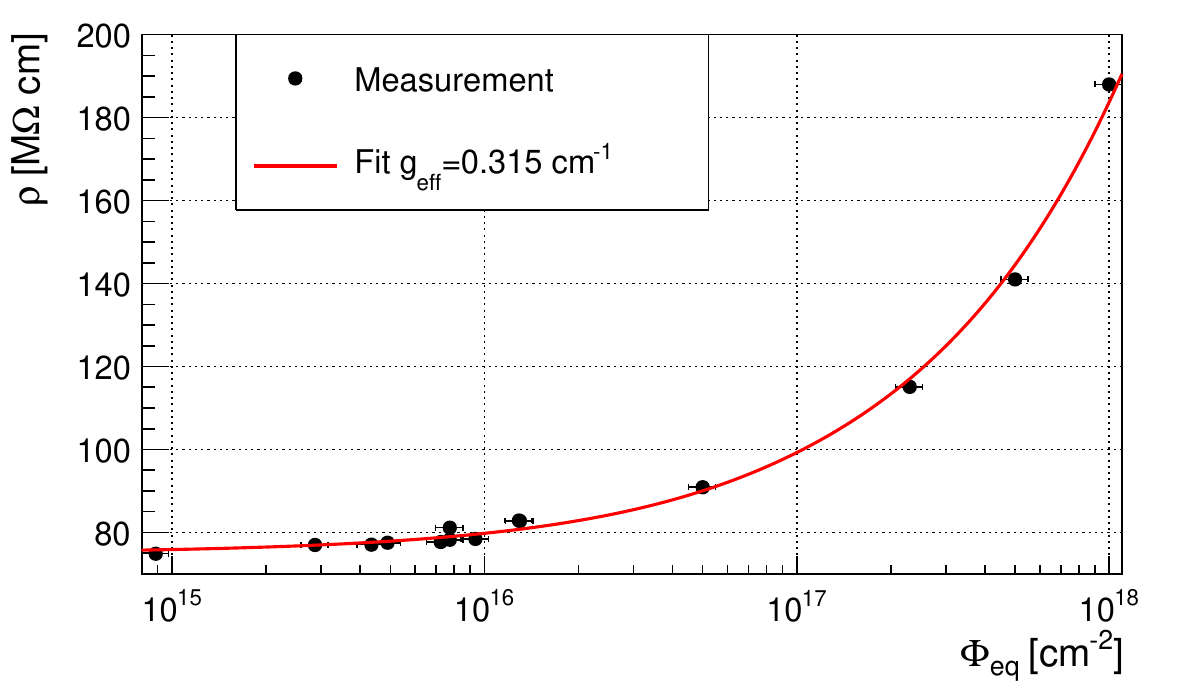}
\caption{\label{fig:fit} A fit of the measured resistivity at $T=243.15$\,K using the Masetti model for ionized-impurity scattering.
}
\end{figure}

The low-field mobilities were obtained by using the default parameters of the empirical Masetti\,\cite{masetti} model fitted for carrier scattering at Phosphorus (electron mobility $\mu_\mathrm{0}^\mathrm{e}\left(N_{\mathrm{P}}\right)$) and Boron (hole mobility $\mu_\mathrm{0}^\mathrm{h}\left(N_{\mathrm{B}}\right)$) dopants. The ionized defect concentrations $N_\mathrm{P,B}$ were replaced by an effective concentration $N_\mathrm{ion}=g_\mathrm{eff}\cdot\Phi_{\mathrm{eq}}$ with an effective ionized-defect introduction rate $g_\mathrm{eff}$ as the single fit parameter. 
The temperature dependence was introduced by scaling the low-field mobilities $\mu_\mathrm{0}^\mathrm{e,h}(T)=\mu_\mathrm{0}^\mathrm{e,h}(300\text{\,K})\cdot\left(\nicefrac{T}{300\text{\,K}}\right)^{\alpha_\mathrm{e,h}}$ with the parameters $\alpha_\mathrm{e}=2.5$ and $\alpha_\mathrm{h}=2.2$ of Ref.\,\cite{canali}. 
The measured constant resistivity at low bias voltages was fitted using
\begin{equation}
\rho\left(\Phi_{\mathrm{eq}},T\right) = c\left[e_\mathrm{0} n_\mathrm{i}\left(T\right)\left( \sqrt{\frac{n}{p}} \cdot \mu_\mathrm{0}^\mathrm{e}\left(\Phi_{\mathrm{eq}}, T\right) + \sqrt{\frac{p}{n}} \cdot \mu_\mathrm{0}^\mathrm{h}\left(\Phi_{\mathrm{eq}}, T\right) \right)\right]^{-1}
\end{equation}
with $\sqrt{\nicefrac{p}{n}}=2.00$ independent of $\Phi_{\mathrm{eq}}$ from the HPTM TCAD simulations, $n_\mathrm{i}(T)$ of Ref.\,\cite{ni} with the band gap $E_\mathrm{g}(T)$ of Slotboom\,\cite{slotboom}, the elementary charge $e_0$, and a constant scaling factor $c=1.077$ which corrects uncertainties of the diode thickness, effective pad area, and the parameterization for $n_\mathrm{i}$. All data of Figure\,\ref{fig:fit} and the data between 230\,K and 260\,K of Figure\,\ref{fig:tempdep} was included in the fit.

Figure\,\ref{fig:fit} shows the fit of the data as a function of the fluence at $T=243.15$\,K. The data for $\Phi_{\mathrm{eq}} \geq 5 \cdot 10^{16}\,$cm$^{-2}$ is new data of the neutron-irradiated samples of this work while the data for $\Phi_{\mathrm{eq}} \leq 1.3 \cdot 10^{16}\,$cm$^{-2}$ was measured for similar proton-irradiated samples in Ref.\,\cite{thesis}. The fit model describes the data very well for an effective ionized-defect introduction rate of $g_\mathrm{eff}=0.315$\,cm$^{-1}$.

\begin{figure}[htbp]
\centering 
\begin{subfigure}{0.48\textwidth}
  \centering
  \includegraphics[width=\textwidth]{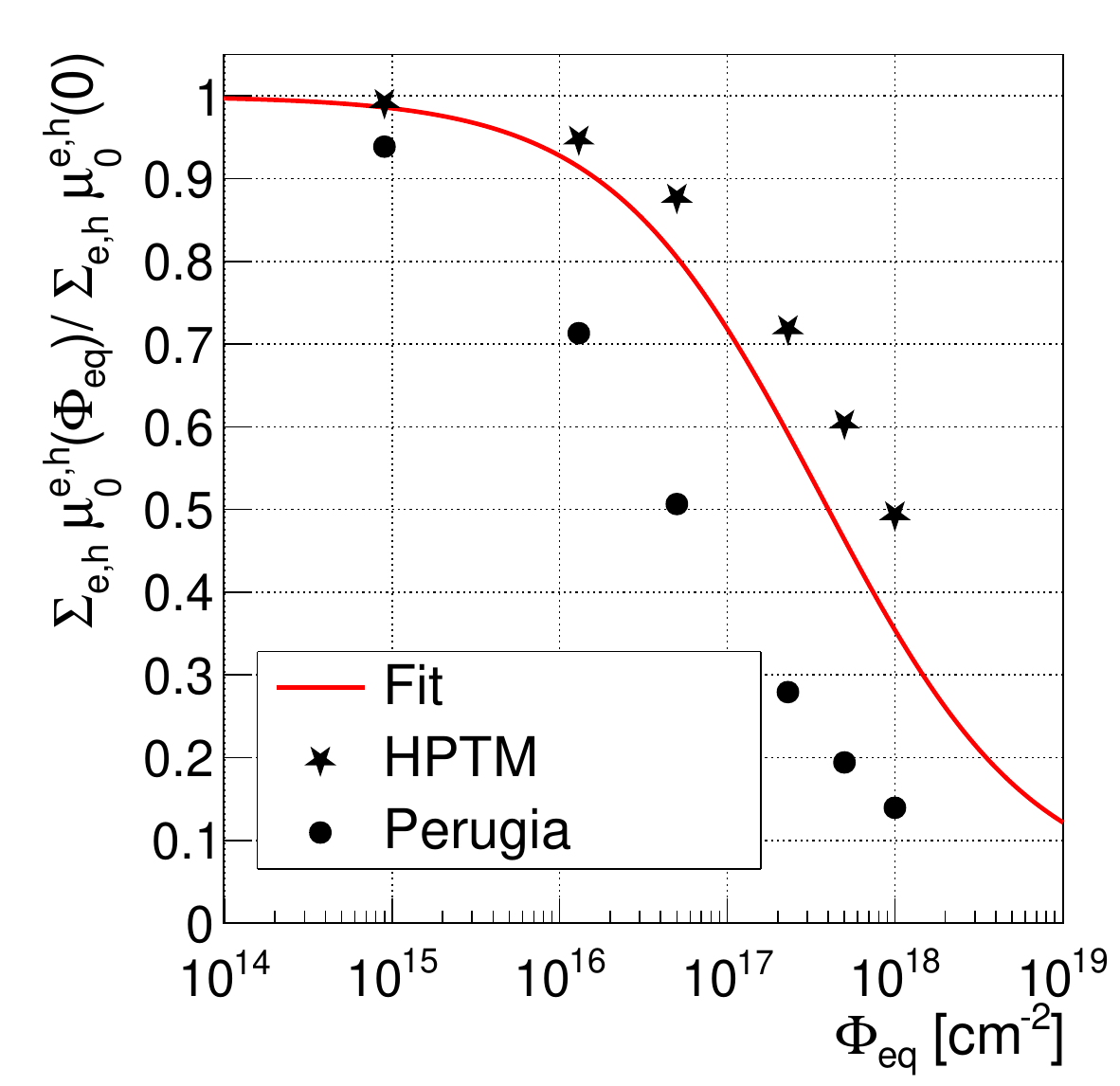}
  \caption{Relative change of $\mu_\mathrm{0}^\mathrm{e}+\mu_\mathrm{0}^\mathrm{h}$.}
  \label{fig:mu_xtr}
\end{subfigure}\hfill
\begin{subfigure}{0.48\textwidth}
  \centering
  \includegraphics[width=\textwidth]{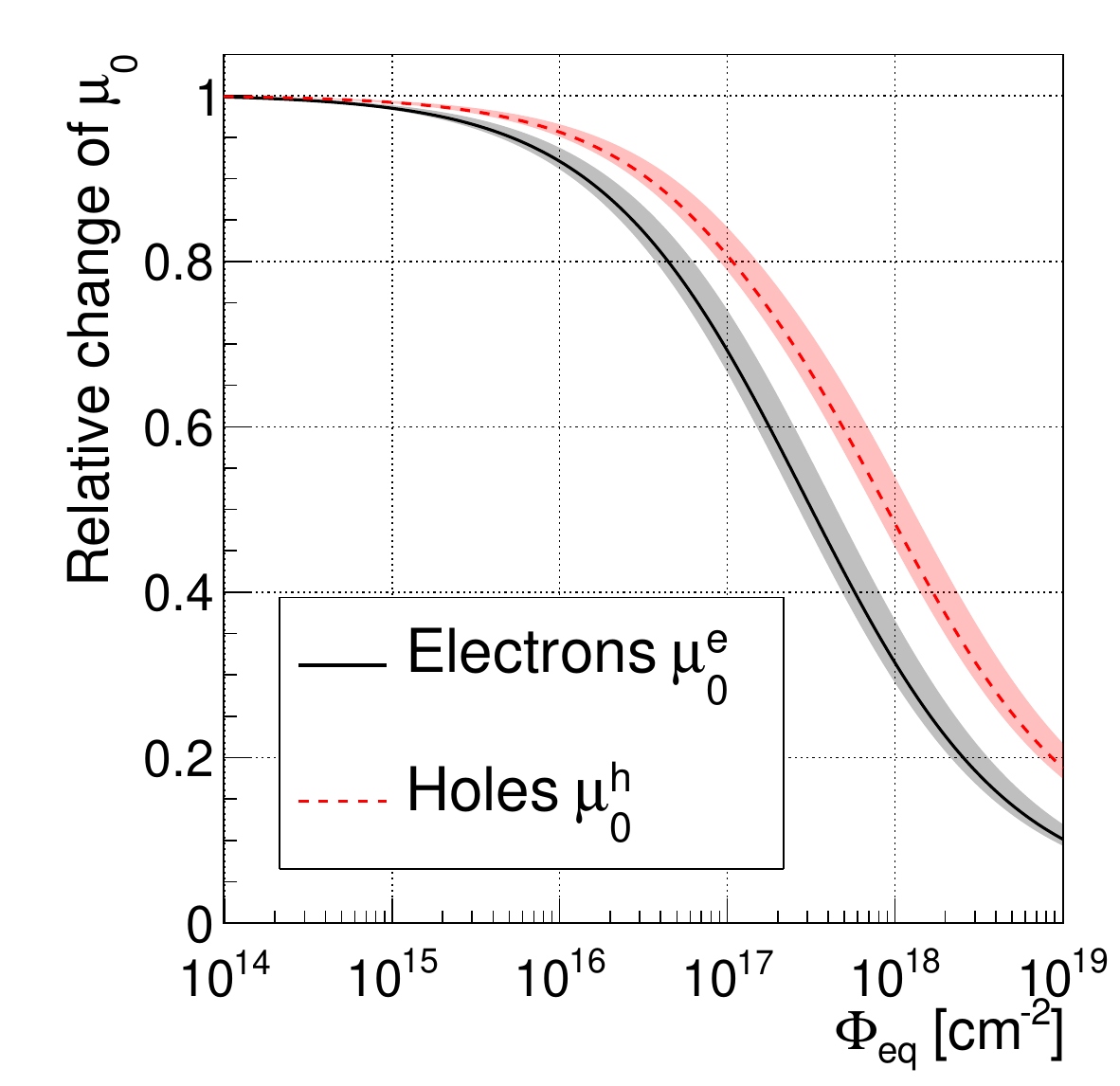}
  \caption{Relative change of $\mu_\mathrm{0}^\mathrm{e,h}$ with uncertainties.}
  \label{fig:mu_err}
\end{subfigure}
\caption{\label{fig:mobs} (a) Predicted change of the sum of the low-field mobilities vs. fluence for the fitted model of this work (line) compared to the simulations described in the text. (b) Change of the low-field mobilities as a function of the fluence with uncertainties. Both figures are plotted for $T=243.15$\,K.
}
\end{figure}

Figure\,\ref{fig:mu_xtr} shows a comparison of the fit results to the simulations. The Perugia simulation overestimates the reduction of the mobilities, probably because it also includes non-ionized defects for the calculation of the mobility. The HPTM simulation underestimates the reduction. However, the HPTM parameters were obtained without taking the reduction of the mobility into account and the model was not yet adapted for extreme fluences. Some previous measurements of the Hall mobility and other methods suggested a much stronger decrease of the mobilities with the fluence\,\cite{hallrhomob,vaitkusmob}, not compatible with other literature\,\cite{fretwurst} and the measurements and simulations presented here. However, the results presented here might be slightly underestimated since band-gap narrowing was not taken into account in the fit, which might become relevant for ionized-impurity concentrations $>10^{17}$\,cm$^{-3}$, but was not observed\,\cite{klanner} for fluences $\leq10^{17}$\,cm$^{-2}$.

Figure\,\ref{fig:mu_err} shows the relative reduction of the low-field mobilities of electrons and holes separately as a function of fluence. In the Masetti model the electron mobility reduces earlier than the hole mobility. The uncertainty bands in the figure cover $1<\nicefrac{p}{n}<8$ to account for a potential inaccuracy of the Fermi level simulated with the HPTM defects and since the simulated $\nicefrac{p}{n}$ exhibited a slight temperature dependence with a change of $\pm10$\% between 230\,K and 260\,K.

\subsection*{Temperature dependence}

\begin{figure}[htbp]
\centering 
\includegraphics[width=0.8\textwidth ]{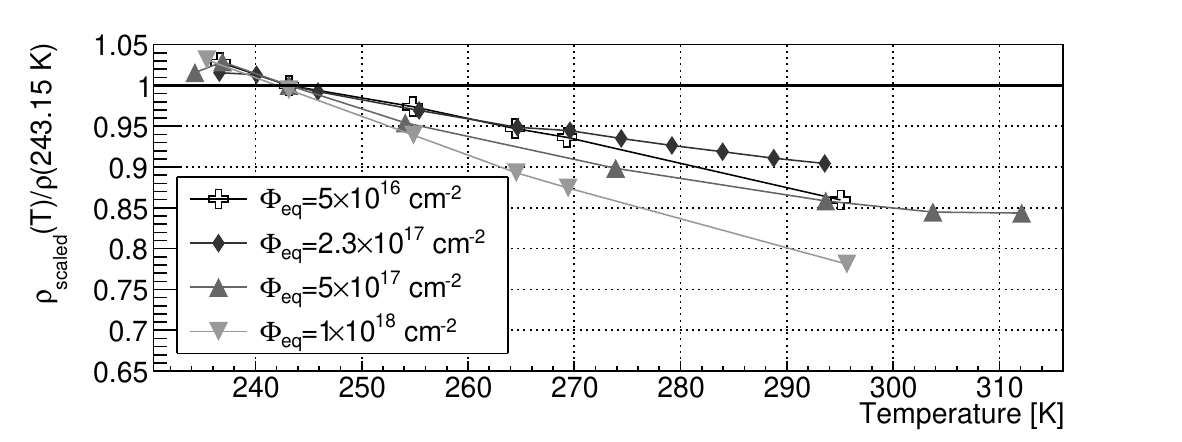}
\caption{\label{fig:tempdep} Measured ohmic resistivity of diodes irradiated to different fluences $\Phi_{\mathrm{eq}}$ at different temperatures, scaled to $T=243.15$\,K using the model described in the text.
}
\end{figure}


Figure\,\ref{fig:tempdep} shows the measured resistivities for different fluences and temperature scaled to 243.15\,K. In the fit range between 230\,K and 260\,K the deviation is less than 10\%. For higher temperatures the temperature scaling does not work very well. This is not surprising since the temperature scaling used for the Masetti model does not take into account that the ionized-impurity scattering mobility $\mu_\mathrm{I}$ and the lattice scattering mobility have opposite temperature dependencies\,\cite{longmyers}. So far, it was not possible to fit the ionized-impurity scattering mobilities separately using the expected temperature dependence $\mu_\mathrm{I}(T) \propto T^{3/2}$ and Matthiessen's rule because of the strong correlation of the fit parameters.

\section{Summary}

The reduction of the low-field mobilities in silicon as a function of the particle fluence was determined by fitting an existing model for ionized-impurity scattering to measured resistivities of diodes irradiated to extreme fluences up to $\Phi_{\mathrm{eq}} =10^{18}\,$cm$^{-2}$. The reduction of the mobilities becomes noticeable for $\Phi_{\mathrm{eq}} > 10^{15}\,$cm$^{-2}$ and at $\Phi_{\mathrm{eq}} = 10^{17}\,$cm$^{-2}$ the mobilities are reduced by about 30\%. The results are somewhat consistent with TCAD simulations using radiation-damage models developed for lower fluences. 
The model presented here uses a fitted effective ionized-defect introduction rate of $g_\mathrm{eff}=0.315$\,cm$^{-1}$ as the single parameter to describe the mobilities as a function of the fluence.

\acknowledgments

This research was supported by the German Federal Ministry of Education and Research (BMBF) as part of Verbundprojekt 05H2024. The authors would like to thank the crew at the TRIGA reactor in Ljubljana for help with irradiations. The authors acknowledge the financial support from the Slovenian Research and Innovation Agency (research core funding No. P1-0135). This project has received funding from the European Union’s Horizon Europe Research and Innovation programme under Grant Agreement No. 101057511 (EURO-LABS).

\vspace{1em}
\noindent {\large\textbf{Copyright}}

\vspace{0.8em}
2026 CERN for the benefit of the ATLAS ITk Collaboration. CC-BY-4.0 license.

\end{document}